\documentclass[submission, Phys]{SciPost}
\usepackage[utf8]{inputenc}

\usepackage[bbgreekl]{mathbbol}
\usepackage{amsfonts}
\DeclareSymbolFontAlphabet{\mathbb}{AMSb}
\DeclareSymbolFontAlphabet{\mathbbl}{bbold}
\usepackage{bbm}
\usepackage{gensymb,amssymb,graphicx,color,amsmath,mathtools,bm}
\usepackage{braket}
\usepackage[normalem]{ulem}
\usepackage[caption=false]{subfig} 

\newcommand{\secref}[1]{Sec.\,\ref{#1}}
\newcommand{\refcite}[1]{Ref.\,\cite{#1}}

\newcommand{\eqnref}[1]{Eq.\,\eqref{#1}}
\newcommand{\eqsref}[1]{Eqs.\,\eqref{#1}}
\newcommand{\figref}[1]{Fig.\,\ref{#1}}
\newcommand{\figsref}[1]{Figs.\,\ref{#1}}
\newcommand{\sfigref}[2]{Fig.\,\hyperref[#1]{\ref{#1}#2}}

\DeclareMathOperator{\Tr}{Tr}
\DeclareMathOperator{\RE}{Re}

\newcommand{\appref}[1]{Appendix\,\ref{#1}}

\definecolor{kspink}{RGB}{200,0,200}

\hypersetup{
  colorlinks = true,
  urlcolor = blue,
  pdfauthor = {Sayak Guha Roy Kevin Slagle},
  pdftitle = {Guha Roy and Slagle - 2023 - Interpolating Between the Gauge and Schrodinger Pictures of Quantum Dynamics}
}

\begin{document}

\begin{center}{\Large \textbf{
Interpolating Between the Gauge and Schr\"odinger Pictures of Quantum Dynamics
}}\end{center}

\begin{center}
Sayak Guha Roy\textsuperscript{1}*,
Kevin Slagle\textsuperscript{2}$^\dagger$
\end{center}

\begin{center}
{\bf 1} Department of Physics and Astronomy, Rice University, Houston, Texas 77005, USA
\\
{\bf 2} Department of Electrical and Computer Engineering, Rice University, Houston, Texas 77005 USA
\\
* sayak.guha.roy@rice.edu \\
$\dagger$ kevin.slagle@rice.edu
\end{center}

\begin{center}
\today
\end{center}


\section*{Abstract}
{\bf
Although spatial locality is explicit in the Heisenberg picture of quantum dynamics, spatial locality is not explicit in the Schr\"odinger picture equations of motion. The gauge picture is a modification of Schr\"odinger's picture such that locality is explicit in the equations of motion. In order to achieve this explicit locality, the gauge picture utilizes (1) a distinct wavefunction associated with each patch of space, and (2) time-dependent unitary connections to relate the Hilbert spaces associated with nearby patches. In this work, we show that by adding an additional spatially-local term to the gauge picture equations of motion, we can effectively interpolate between the gauge and Schr\"odinger pictures, such that when this additional term has a large coefficient, all of the gauge picture wavefunctions approach the Schr\"odginer picture wavefunction (and the connections approach the identity).
}

{
\hypersetup{linkcolor=black}
\vspace{10pt}
\noindent\rule{\textwidth}{1pt}
\tableofcontents\thispagestyle{fancy}
\noindent\rule{\textwidth}{1pt}
\vspace{10pt}
}

\section{Introduction}
\label{sec:intro}

The dynamics of a physical system is \emph{explicitly spatially local} if
  the degrees of freedom are local (i.e. can be associated with a position in space)
  and if the time dependence of the degrees of freedom only depend on sufficiently nearby degrees of freedom.
In the Schr\"odinger picture of quantum dynamics,
  the wavefunction is the only time-dependent degree of freedom.
But the wavefunction is not a local degree of freedom;
  it is a global degree of freedom since it can not be associated with a particular position in space.
As such, the Schr\"odinger picture does not exhibit explicit locality.
In contrast, the Heisenberg picture does exhibit explicit locality \cite{Heisenberg-locality,Deutsch_2011}
  since the time dependence of local operators only depends on nearby local operators (for local Hamiltonians).

\begin{figure}
    \centering
    \includegraphics{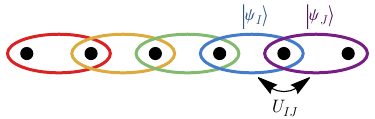}
    \caption{An example of a collections of patches (colored ovals) consisting of pairs of neighboring qubits (black dots).
    A local wavefunction $\ket{\psi_I}$ is associated with each patch $I$, and the Hilbert spaces associated with neighboring patches are related by unitary connections $U_{IJ}$.}
    \label{fig:patches}
\end{figure}

Since locality is of fundamental importance to theoretical physics,
  a modified version of the Schr\"odinger picture \cite{slagle2022gauge} was recently developed such that locality is explicit in the equations of motion.
To formulate this new picture, we first choose a set of local patches (indexed by capital letters $I$, $J$, or $K$) of space (or the lattice),
  as depicted in \figref{fig:patches}.
In the simplest setting, the patches can be taken to be the spatial support of the different Hamiltonian terms.
A distinct \emph{local wavefunction} $\ket{\psi_J}$ is associated with each patch $J$.
Furthermore, the Hilbert spaces of nearby patches ($I$ and $J$) are related by time-dependent unitary transformations $U_{IJ}$.
These unitary transformations resemble gauge connections in a lattice gauge theory (while the local wavefunctions resemble Higgs fields \cite{slagle2022quantum}),
  which motivates the name ``gauge picture'' for this picture of quantum dynamics.

The equations of motion in the gauge picture are given by
\begin{equation}
\begin{aligned}
    \partial_t\ket{\psi_I} &= -iH_{\left\langle I \right\rangle}^\text{G}\ket{\psi_I} \\
    \partial_tU_{IJ} &= -iH_{\left\langle I \right\rangle}^\text{G} U_{IJ} + i U_{IJ}H_{\left\langle J \right\rangle}^\text{G}
\end{aligned} \label{eq:gaugeEoM}
\end{equation}
$H_{\left\langle I \right\rangle}^\text{G}$ is the sum of the Hamiltonian terms on all patches that overlap with patch $I$
\begin{equation}
    H_{\left\langle I \right\rangle}^\text{G} = \sum_J^{J\cap I \neq 0} U_{IJ} H_J^\text{G} U_{JI}
    \label{eq:gaugeH}
\end{equation}
$H_J$ is a Hamiltonian term supported on patch $J$,
  such that the Hamiltonian of the entire system is $H= \sum_J H_J$.
We use S, H, and G superscripts to distinguish time-dependent operators in the Schr\"odinger, Heisenberg, and gauge pictures respectively.
Time-independent operators in the Schr\"odinger picture are also time-independent in the gauge picture.

The local wavefunctions $\ket{\psi_I}$ are local in the sense that their time dynamics only depends on nearby degrees of freedom (i.e. $\ket{\psi_I}$ and $U_{IJ}$ where $I$ and $J$ overlap).
As a consequence, although $\ket{\psi_I}$ lives in an exponentially large Hilbert space for a large many-body system
  (e.g. of dimension $2^n$ for a system of $n$ qubits),
  by itself, $\ket{\psi_I}$ only encodes enough information to compute expectation values of operators supported on the patch $I$.
The information that describes operators outside patch $I$ is typically scrambled,
  and one must use a string of connections $U_{IJ}$, e.g. $\braket{\psi_I | A_I^\text{G} U_{IJ} U_{JK} B_K^\text{G} | \psi_K}$,
  to unscramble this information to compute long-range correlation functions.

In this work,
  we question to what extent it is possible to obtain explicitly local equations of motion (such as the gauge picture)
  such that distant information is not scrambled in this way.
That is,
  we ask if it is possible to modify the gauge picture such that local wavefunctions are approximately equal to the Schr\"odinger picture wavefunction: $\ket{\psi_I} \approx \ket{\psi}$.
To achieve this,
  we consider adding a local term to the equations of motion that drives the connections $U_{IJ}$ towards the identity
  (without affecting any expectations values or operator time-dependence in the gauge picture).
We show that if this new term has a large coefficient $\gamma$,
  then the connections are approximately equal to the identity
  and all of the local wavefunctions in the gauge picture are approximately equal to the Schr\"odginer picture wavefunction.
In this sense, this coefficient is capable of interpolating between the gauge and Schr\"odinger pictures.
However, we find that the magnitude of the $\gamma$ coefficient must scale exponentially with system size in order to keep the deviation between the two wavefunctions below a constant bound.

In \secref{sec:gaugePicture}, we briefly review a derivation of the gauge picture of quantum dynamics.
With the derivation fresh in our mind,
  it is clear what kinds of modifications can be straightforwardly made to the gauge picture.
In \secref{sec:deriveEoM},
  we derive an additional term, with coefficient $\gamma$,
  that we can add to the gauge picture in order to
  interpolate between the gauge and Schr\"odinger pictures.
In \secref{sec:scaling}, 
  we estimate how much the modified gauge picture will deviate from Schr\"odinger's picture (i.e. how much $U_{IJ}$ deviates from the identity)
  in the limit of large $\gamma$.
In Sections~\ref{sec:simulations} and \ref{sec:app_a},
  we validate our estimates using numerical simulations of the 1D transverse-field Ising model \cite{PFEUTY197079,PhysRevLett.25.443}
  in a longitudinal field \cite{PhysRevE.99.012122}.

\section{Review of the gauge picture}
\label{sec:gaugePicture}

We wish to modify the gauge picture such that the local wavefunctions in the gauge picture are approximately equal to the Schr\"odinger picture wavefunction.
At the same time, we want the modified gauge picture to be an exact description of the quantum dynamics, while still retaining the explicit locality that originally motivated the gauge picture.
In order to deduce the ideal modification,
  it is useful to review how the gauge picture is derived.

The gauge picture can be derived from the Heisenberg picture,
  which also features explicitly local equations of motion.
Consider a local Hamiltonian
\begin{equation}
  H = \sum_J H_J \label{eq:sumHJ}
\end{equation}
that is a sum over Hamiltonian terms $H_J$, each supported on some patch $J$ of the lattice.
A local operator $A_I$ supported on a patch ($I$) is time-evolved in the Heisenberg picture via
\begin{align}
  \partial_t A_I^\text{H} &= i [H^\text{H}, A_I^\text{H}] \nonumber\\
  &= i [H_{\braket{I}}^\text{H}, A_I^\text{H}] \label{eq:Heisenberg}
\end{align}
For simplicity, we assume that operators have no explicit time dependence.
We use S, H, and G superscripts to distinguish time-dependent operators in the Schr\"odinger, Heisenberg, and gauge pictures. 
In the second line above,
  we note that most terms in the Hamiltonian cancel out in the commutator due to locality,
  and only the following Hamiltonian terms contribute:
\begin{equation}
    H_{\left\langle I \right\rangle}^\text{H} = \sum_J^{J\cap I \neq 0} H_J^\text{H}
    \label{eq:HI}
\end{equation}
  where $\sum_J^{J\cap I \neq 0}$ denotes a sum over all patches that overlap with patch $I$.

Therefore, the Heisenberg picture equation of motion \eqref{eq:Heisenberg} is explicitly local \cite{Heisenberg-locality,Deutsch_2011}.
In order to obtain the gauge picture,
  we need to push the time dependence from the operators into the wavefunction.
This is achieved using the following unitary mapping:
\begin{equation}
\begin{aligned}
    \ket{\psi_I} = U_I \ket{\psi^\text{H}}\\
    A_I^\text{G}=U_IA_I^\text{H}U_I^{\dagger}
\end{aligned}  \label{eq:gaugeHeisenberg}
\end{equation}
Similar to how the Schr\"odinger and Heisenberg picture wavefunction and operators are related by a unitary transformation,
  the above equation relates the wavefunction and operators in the Heisenberg picture (right hand side)
  to those in the gauge picture (left hand side)
  using a collection of unitary transformations $U_I$.
An important difference, however,
  is that in order to maintain a sense of local dynamics for the wavefunction,
  we must use a different unitary transformation for each patch of space,
  which results in the \emph{local wavefunctions} $\ket{\psi_I}$.

Since $U_I$ is unitary, the time derivative of $U_I$ can be expressed in terms of a unitary operator $G_I(t)$:
\begin{equation}
    \partial_t U_I = -iG_I U_I \label{eq:dUI}
\end{equation}
Plugging \eqsref{eq:gaugeHeisenberg} and \eqref{eq:dUI} into $\partial_t\ket{\psi^\text{H}}=0$ and the local Heisenberg equation \eqref{eq:Heisenberg} of motion yields:
\begin{equation}
\begin{aligned}
    \partial_t \ket{\psi_I} &= -iG_I\ket{\psi_I}\\
    \partial_tA_I^\text{G} &= i[H_{\left\langle I \right\rangle}^\text{G} - G_I,A_I^\text{G}] 
\end{aligned} \label{eq:gaugeEoMG}
\end{equation}
In the gauge picture, $H_{\langle I \rangle}$ from \eqnref{eq:HI} takes a modified form:
\begin{align}
    H_{\left\langle I \right\rangle}^\text{G}
      &= U_I H_{\left\langle I \right\rangle}^\text{H} U_I^\dagger \nonumber\\
      &= \sum_J^{J\cap I \neq 0} U_{IJ} H_J^\text{G} U_{JI}
    \label{eq:HIG}
\end{align}
Above, we have defined the \emph{connections}
\begin{equation}
  U_{IJ} = U_I U_J^{\dagger}
\end{equation}
From \eqnref{eq:dUI}, we see that the connections evolve as
\begin{equation}
    \partial_t U_{IJ} = -i G_I U_{IJ} + U_{IJ} G_J \label{eq:dUIJ}
\end{equation}

$U_{IJ}$ connects the wavefunctions of different patches via
\begin{equation}
    U_{IJ}\ket{\psi_J} = \ket{\psi_I}
\end{equation}
  which follows from \eqnref{eq:gaugeHeisenberg}.
The unitary connections $U_{IJ}$ between different patches $(I,J,K)$ are analogous to ``flat'' gauge fields; i.e. they obey
\begin{equation}
    U_{IJ}U_{JK}=U_{IK}
\end{equation}

In the gauge picture \cite{slagle2022gauge}, 
  we choose 
\begin{equation}
\begin{aligned}
    G_I &= H_{\left\langle I \right\rangle}^\text{G} \\
    U_I(t=0) &= \mathbbm{1}
\end{aligned}
\end{equation}
  so that local operators have no time dependence in \eqnref{eq:gaugeEoMG}
  and are equal local operators in the Schr\"odinger picture.
This leads to the gauge picture equations of motion in \eqnref{eq:gaugeEoM}.

\section{Modified Gauge Picture}
\label{sec:deriveEoM}

In this section,
  we derive how the gauge picture can be modified such that the connections can be kept close to the identity.
From the previous section,
  we see that any choice of Hermitian $G_I$ leads to valid equations of motion.
Let us decompose $G_I$ as
\begin{equation}
    G_I = H_{\left\langle I \right\rangle}^\text{G} + \gamma X_I
    \label{eq:G_int}
\end{equation}
  where $\gamma$ is a real-valued constant and $X_I$ is a Hermitian operator.
If $X_I$ commutes with local operators $A_I$ that only act on patch $I$,
  then local operators will still be time-independent in \eqnref{eq:gaugeEoMG}.
In this section,
  we will derive a choice of $X_I$ such that the connections are pushed towards the identity.

We can quantify how close a connection $U_{IJ}$ is to the identity via its trace, $\Tr U_{IJ}$,
  which increases as $U_{IJ}$ approaches the identity.
To be precise, we define
\begin{equation}
    S_{IJ}(t) = 1-\frac{\RE \Tr (U_{IJ})}{N}
    \label{eq:Sch-ness}
\end{equation}
where $N$ is the Hilbert space dimension
  (e.g. $N=2^n$ for a system with $n$ qubits).
The value of $S_{IJ}(t)$ ranges between $0$ and $2$,
  with $S_{IJ}(t) = 0$ when $U_{IJ}$ is the identity.

Therefore, we want to choose $X_I$ such that the average $\partial_t S_{IJ}(t)$ is minimized (while holding a norm of $X_I$ fixed).
The averaged time derivative is
\begin{equation}\begin{aligned}
    \partial_t \sum_{IJ}^{I \cap J \neq \emptyset} S_{IJ}(t) 
    &= -\frac{1}{N} \sum_{IJ}^{I \cap J \neq \emptyset} \Tr \partial_t U_{IJ} \\
    &= -\frac{1}{N} \sum_{IJ}^{I \cap J \neq \emptyset} \Tr (-i G_I U_{IJ} + i U_{IJ} G_J) \\
    &= -\frac{1}{N} \sum_I \Tr G_I \underbrace{\sum_J^{I\cap J \neq \emptyset} \left(-i U_{IJ} + i U_{IJ}^\dagger \right)}_{\tilde{X}_I}
\end{aligned}\end{equation}
$\sum_{IJ}^{I \cap J \neq \emptyset}$ sums over all patches $I$ and $J$ that have nonzero overlap.
In the last line, we identify a candidate
\begin{equation}
    \Tilde{X}_I = \sum_J^{I\cap J \neq \emptyset} (-i) \left(U_{IJ} - U_{IJ}^\dagger \right) \label{eq:tildeX}
\end{equation}
for $X_I$,
  which will contribute a negative derivative in the total $S_{IJ}(t)$.
But as previously mentioned,
  we want $X_I$ to commute with any local operator $A_I$ supported on a patch $I$
  so that local operators are time-independent.
To achieve this, we simply define $X_I$ as $\Tilde{X}_I$ after taking the partial trace over qubits in patch $I$:
\begin{equation}
    X_I = \Tr_I \Tilde{X}_I
    \label{eq:X_I}
\end{equation}
Therefore with this choice of $X_I$ and for large $\gamma$,
  the $\gamma X_I$ term in \eqnref{eq:G_int} can be expected to drive
  $\sum_{IJ}^{I \cap J \neq \emptyset} S_{IJ}(t)$ towards zero,
  therefore pushing  the connections $U_{IJ}$ toward identity.

\section{Scaling Hypothesis}
\label{sec:scaling}

We can estimate how effective the $\gamma$ term is at driving the connections towards the identity.
In the previous section,
  we argued that for very large $\gamma$,
    the connections should be very close to the identity: $U_{IJ} \approx \mathbbm{1}$.
We therefore expect the following perturbative expansion in small $\gamma^{-1}$:
\begin{equation}
        \ln U_{IJ} = i \sum_{k=1}^{\infty} \gamma^{-k} A_{IJ}^{(k)}
\label{eq:pertSG}
\end{equation}
  where $A_{IJ}^{(k)}(t)$ are time-dependent Hermitian operators with no dependence on $\gamma$.
Although this expression is clearly valid for sufficiently small times $t$,
  its validity at long times $t \gg \gamma^{-1}$ is not immediately clear.

We previously identified $S_{IJ}(t)$ as a useful metric for how much the connection $U_{IJ}$ deviates from the identity.
If we boldly assume that the above expansion holds at late times,
  then we can can estimate
\begin{align}
    S_{IJ}(t) &= 1- \frac{1}{N} \RE \Tr U_{IJ} \nonumber\\
    &= \gamma^{-2} \frac{\Tr A_{IJ}^{(1)} A_{IJ}^{(1)}}{2N} + O(\gamma^{-3})
\end{align}
  where the terms linear in $i A_{IJ}^{(k)}(t)$ vanish since they have imaginary trace.
We therefore expect that
\begin{equation}
    S_{IJ}(t) \sim \gamma^{-2}
\end{equation}

\section{Simulations}
\label{sec:simulations}

\begin{figure}
    \centering
    \includegraphics[width=7cm]{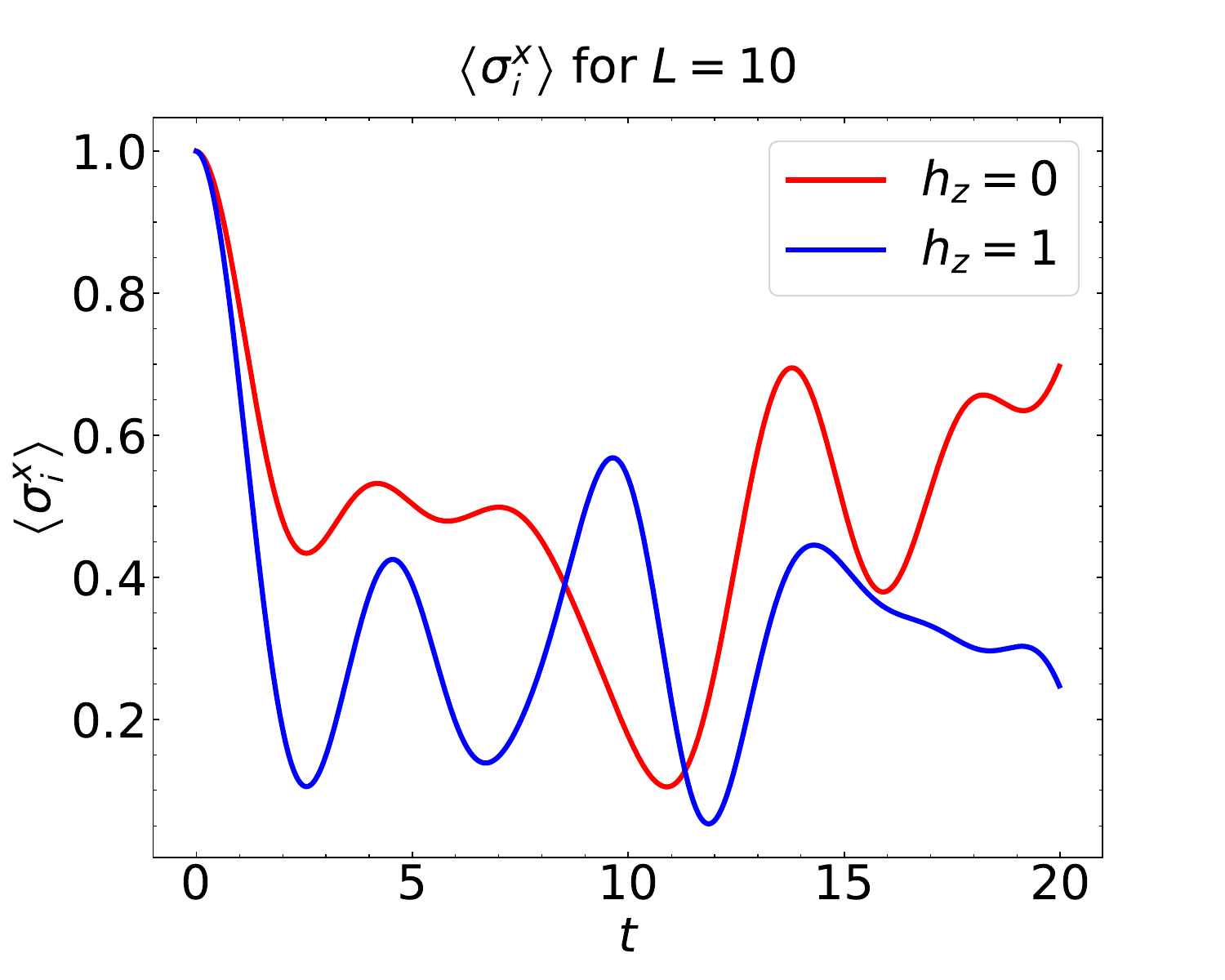}
        \caption{The expectation value $\langle \sigma^x_i \rangle$ vs time $t$ after a quench for a periodic chain of length $L=10$ for $h_\text{z}= 0$ and $h_\text{z}= 1$.}
            \label{fig:X vs t}
\end{figure}

We use numerical simulations to check how effective the $\gamma$ term is driving the connections towards the identity.
We simulate the one-dimensional transverse field Ising model after a quench in two regimes: (1) criticality (and integrable) and (2) with a longitudinal field (which is non-integrable \cite{PhysRevB.92.104306}).
For both cases, we find that for long times $t \gg \gamma^{-1}$,
  large $\gamma$,
  and long chains of length $L$,
  the deviation $S_{IJ}(t)$ from the Schr\"odinger picture
  saturates at a value $S_{IJ}(\infty)$.
We will numerically show that $S_{IJ}(\infty)$ obeys the following scaling in the large $\gamma$ and large $L$ limit:
\begin{equation}
  S_{IJ}(t=\infty) \sim \gamma^{-2} \, e^{a L + b + c/L\cdots}
  \label{eq:fit_asymp}
\end{equation}
  where ``$\cdots$'' denotes subleading terms (and for nearest-neighbor two-qubit patches $I$ and $J$).
Therefore, $S_{IJ}(t=\infty) \sim \gamma^{-2}$ as expected in the previous section.

In the simulations, we initialize the system with all spins pointing in the +X direction such that $\braket{\sigma_i^x}=1$ at $t=0$.
We then consider a time evolution under the transverse field Ising model in a longitudinal field, which has the following Hamiltonian:
\begin{equation}
    H = -J\sum_{\left\langle ij \right\rangle} \sigma^\text{z}_i\sigma^\text{z}_j - h_\text{x} \sum_i \sigma^\text{x}_i - h_\text{z} \sum_i \sigma^\text{z}_i
    \label{eq:hamil}
\end{equation}
$J$ is the Ising interaction strength;
  $\sum_{\left\langle i,j \right\rangle}$ denotes a sum over nearest neighbor sites; 
  $h_\text{x}$ is the transverse field strength;
  $h_\text{z}$ is the longitudinal field strength;
  and $\sigma^{\mu}$ are Pauli operators.
We take $J=h_\text{x}=1$ throughout and consider a periodic chain of length $L$ in two regimes:
  (1) $h_\text{z}=0$, for which the model is critical and integrable (via a mapping to free majorana fermions), and
  (2) $h_\text{z} = 1$, for which the model is gapped and non-integrable \cite{PhysRevB.92.104306}. The time evolution of the expectation value of the spin operator $\sigma^\text{x}$ is shown in \figref{fig:X vs t}.\footnote{%
  Throughout the main text, all simulations are performed with a time step $\delta t = 0.005$ using the modified RK4 Runge Kutta integration method described in Appendix F of \refcite{slagle2022quantum}. The modification is used to maintain $\ket{\psi_I} = U_{IJ} \ket{\psi_J}$ exactly, but has the unfortunate side effect (which is probably preventable) of increasing the integration error from $(\delta t)^4$ to $(\delta t)^3$ at time $t\sim 1$. Nevertheless, we have checked that the time step is sufficiently small to not significantly affect any of our plots.} 

To simulate our modified gauge picture,
  we first choose a set of patches to cover the lattice.
We take the simplest choice of patches $I=\langle i,j \rangle$ that are taken to be nearest-neighbor pairs of qubits $\langle i,j \rangle$, as depicted in \figref{fig:patches}.
Next we must split the Hamiltonian into a sum $H = \sum_I H_I$ of local terms $H_I$:
\begin{equation}
  H_{I=\langle ij \rangle} = -J \sigma^\text{z}_i\sigma^\text{z}_j - \frac{h_\text{x}}{2} (\sigma^\text{x}_i + \sigma^\text{x}_j) - \frac{h_\text{z}}{2} (\sigma^\text{z}_i + \sigma^\text{z}_j)
\end{equation}

Similar to the usual gauge picture, at $t=0$ the local wavefunction are initialized to be equal to the Schr\"odinger picture wavefunction,
  $\ket{\psi_I(t=0)} = \ket{\psi^\text{S}(t=0)}$,
  and the connections are initialized as the identity,
  $U_{IJ}(t=0) = \mathbbm{1}$.
\eqsref{eq:gaugeEoMG} and \eqref{eq:dUIJ},
  with $G_I$ given in \eqnref{eq:G_int},
  are then used to numerically calculate the time-evolved
  local wavefunctions $\ket{\psi_I(t)}$ and connections $U_{IJ}(t)$.
Recall that $G_I$ was chosen such that the local operators (e.g. $\sigma_i^\mu$)
  are equal in the Schr\"odinger and gauge pictures.
Therefore, in order to reduce notational clutter,
  we omit G superscripts and patch index subscripts on the Pauli operators.

\subsection{$h_\text{z}=0$}

\begin{figure}
    \centering
    \subfloat[\label{fig:asymptoteL10}]{\includegraphics[width=7cm]{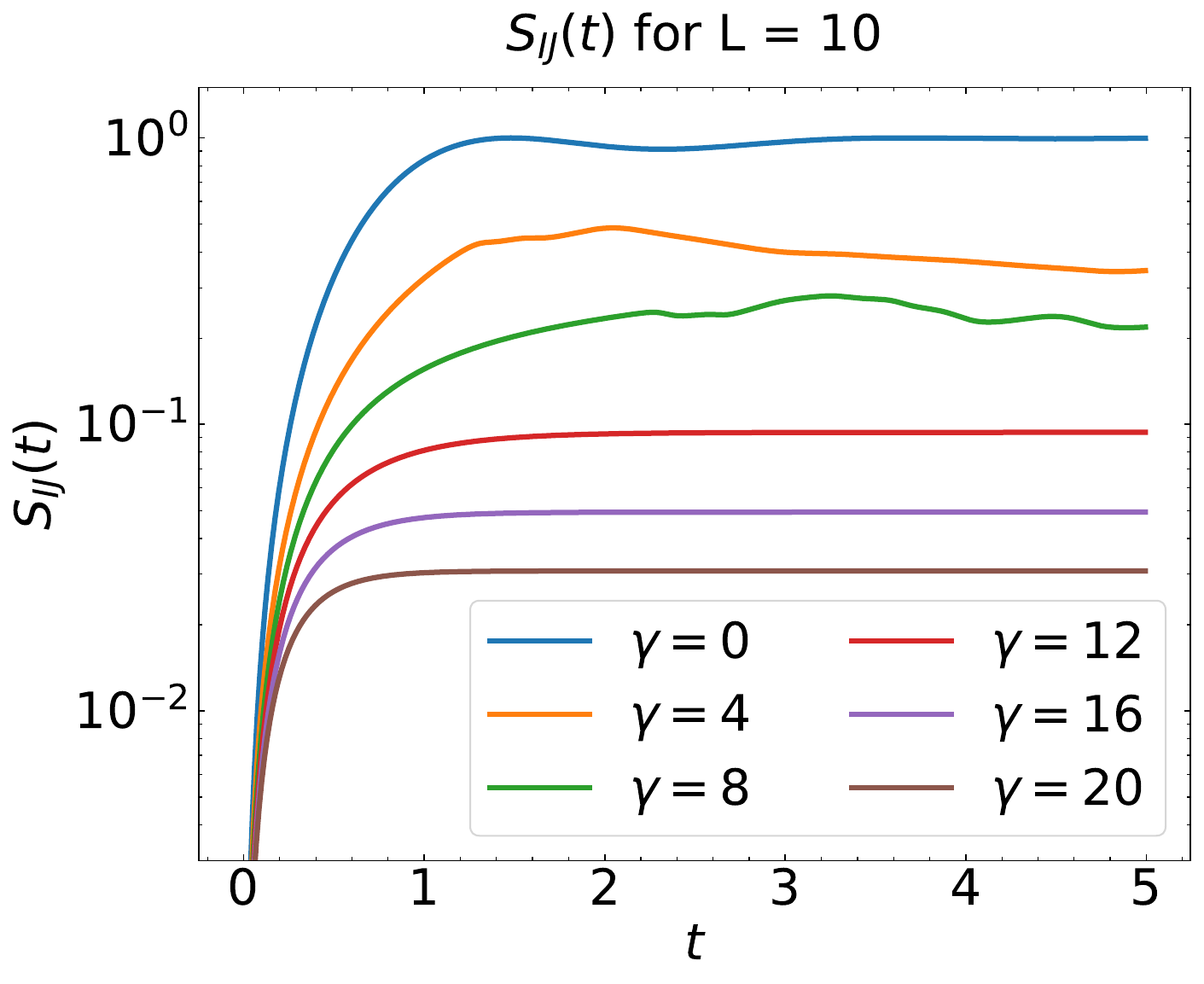}}
    \subfloat[]{\includegraphics[width=7cm]{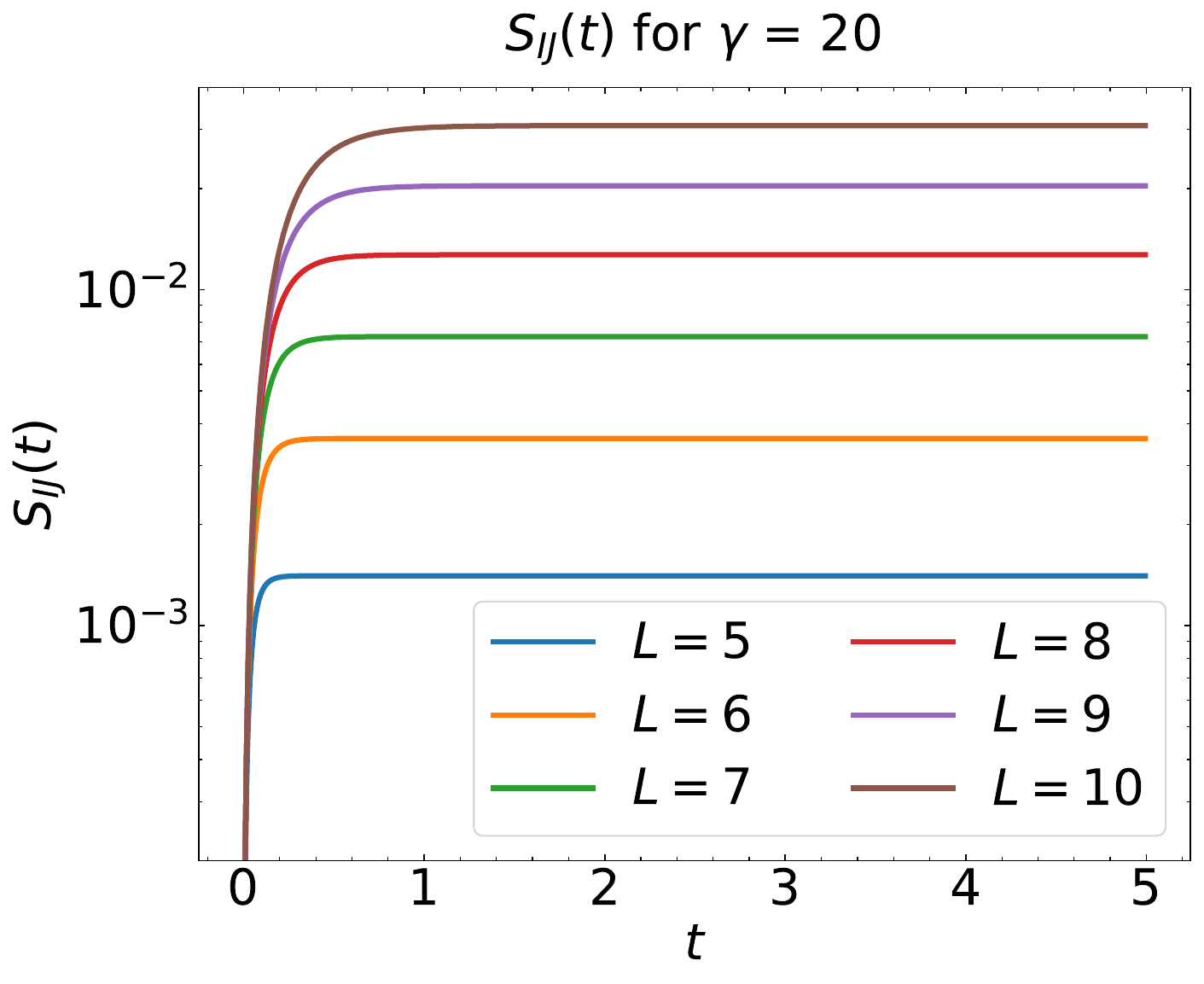}}
    \caption{
    Deviation $S_{IJ}(t)$ of the modified gauge picture from the Schr\"odinger picture
    for {\bf (a)} $L = 10$ and different $\gamma$ and {\bf (b)} for $\gamma = 20$ and different $L$.
    Both plots show data for the critical transverse-field Ising model [\eqnref{eq:hamil}] with $J=h_x=1$ and $h_\text{z}=0$.
    }
    \label{fig:asymptote}
\end{figure}

\begin{figure}
    \centering
    \subfloat[\label{fig:scaling_1a}]{\includegraphics[width=7cm]{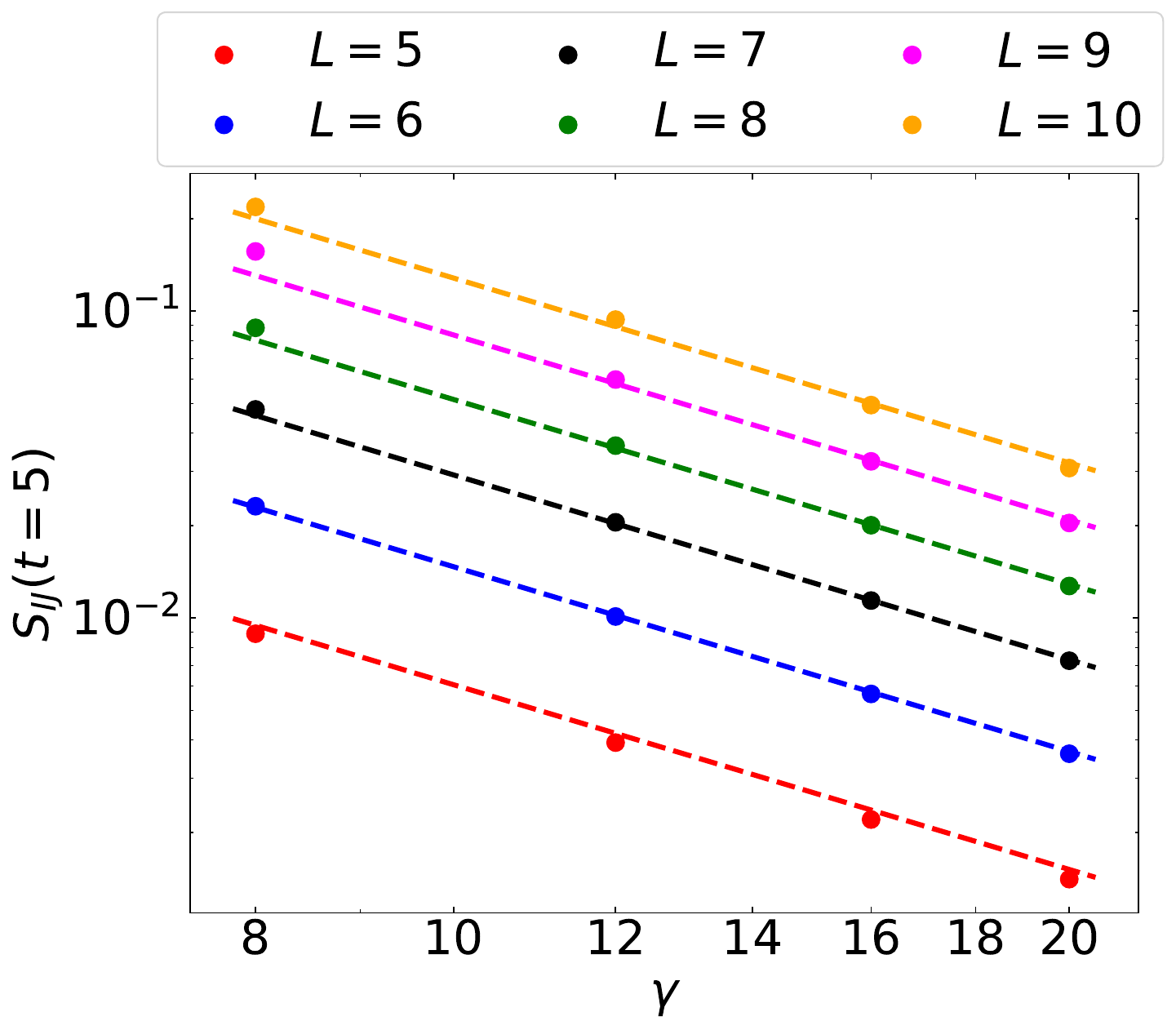}}
    \subfloat[\label{fig:scaling_1b}]{\includegraphics[width=7cm]{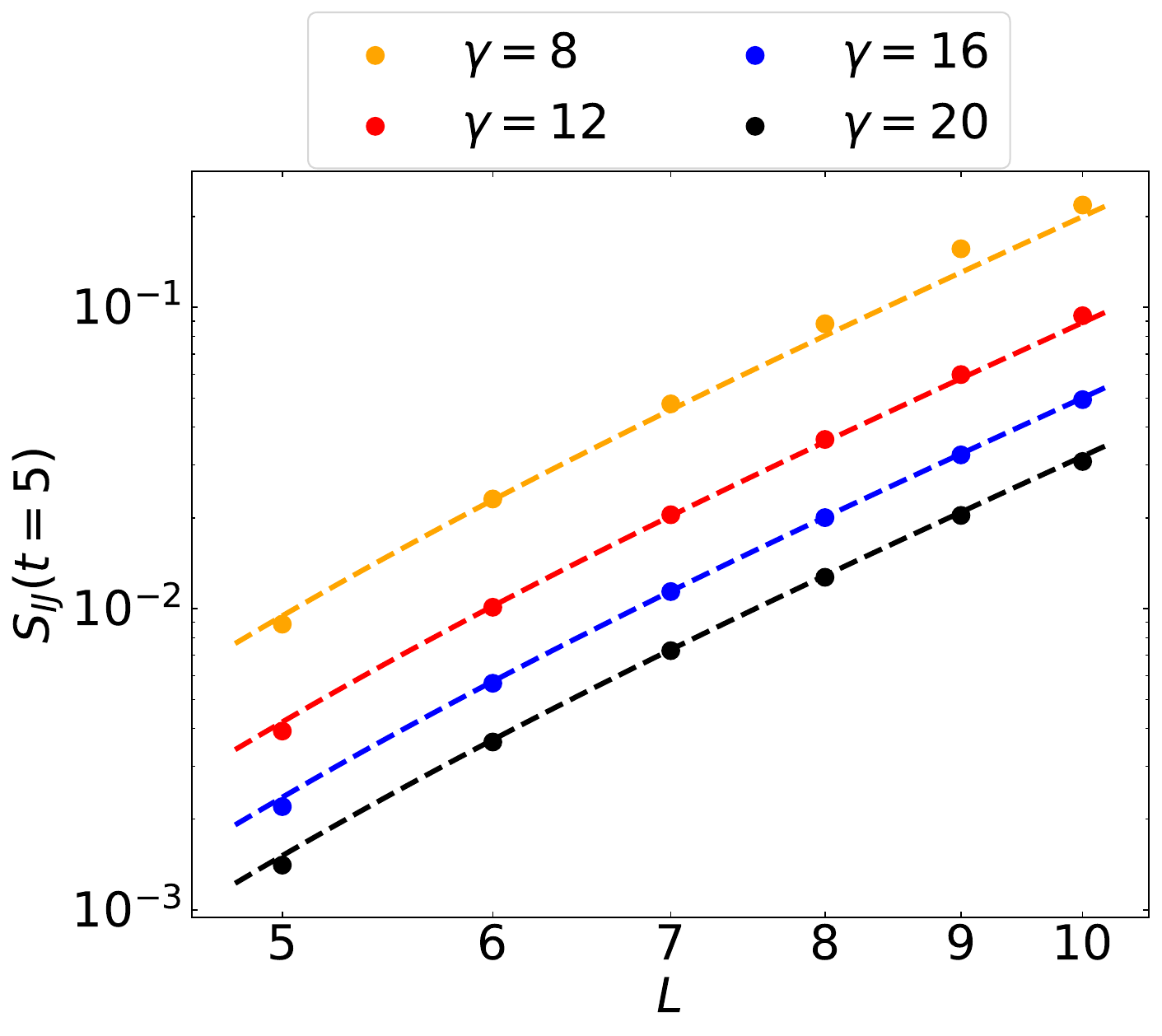}}
    \caption{
    Scaling of the deviation from the Schr\"odinger picture $S_{IJ}(t=5)$ at time $t=5$, which is approximately equal to the asymptote $S_{IJ}(t=\infty)$.
    {\bf (a)} $S_{IJ}(t=5)$ vs $\gamma$ for different $L$.
    {\bf (b)} $S_{IJ}(t=5)$ vs $L$ for different $\gamma$.
    Data is shown for the critical transverse field Ising model model [\eqnref{eq:hamil}] with $J=h_x=1$ and $h_\text{z}=0$.
    }
    \label{fig:scaling}
\end{figure}

Let us consider the $h_\text{z}=0$ case first.
($h_\text{z}=1$ will be qualitatively the same.)
In \figref{fig:asymptote},
we plot $S_{IJ}(t)$ [\eqnref{eq:Sch-ness}] vs time for various $\gamma$ and chain lengths $L$.
We take the patches $I=\langle i-1,i \rangle$ and $J = \langle i,i+1 \rangle$ to be nearest neighbors.
Since the model is translation-symmetric with periodic boundary conditions,
  there is no dependence on $i$ and all $S_{IJ}$ are equal for all nearest-neighbor patches.
Recall that $S_{IJ}(t)$ quantifies how much the modified gauge picture deviates from Schr\"odinger's picture;
  i.e. $S_{IJ}(t)$ quantifies how much the gauge picture connections deviate from the identity.
For sufficiently large $\gamma$, we see that $S_{IJ}(t)$ asymptotes to a constant after time $t \approx 2$,
  and this asymptote decreases with $\gamma$ and increases with system size $L$. 

When $\gamma$ is small, there does not appear to be a clean asymptote to a constant in \figref{fig:asymptoteL10};
  instead, $S_{IJ}(t)$ appears to randomly squiggle around a rough constant.
However, these squiggles appear to be completely absent for $\gamma \geq 12$ in \figref{fig:asymptoteL10}.
Note that the time after which the squiggling begins appears to increase as $\gamma$ increases.
In \secref{sec:app_a}, we provide numerical evidence to show that the time after which the squiggling begins actually diverges at a finite $\gamma$ (which in turn appears to increase with system size). Therefore, for sufficiently large $\gamma$, we expect that there is a clean asymptote to a constant. 

In \figref{fig:scaling}, 
  we study how the $t=\infty$ asymptote of $S_{IJ}(t)$ scales with $\gamma$ and system size $L$.
We fit the data to \eqnref{eq:fit_asymp} and find the following parameters
\begin{align}
  a=2.63 \hspace{5mm} b=0.19 \hspace{5mm}c=-20.63
  \label{eq:param1}
\end{align}
These three parameters where used to simultaneously fit the 10 lines shown in \figref{fig:scaling}.
We only used the twelve 
  data points with $\gamma = 12,16,20$ and $L=7,8,9,10$ to fit the data.
We excluded points with small $\gamma$ that did not have a clean asymptote.
The fit is remarkably clean,
  which is strong evidence for the validity of the scaling shown in \eqnref{eq:fit_asymp}.

\subsection{$h_\text{z}=1$}  

\begin{figure}
    \centering
    \subfloat[\label{fig:asymptoteL10_hz1}]{\includegraphics[width=7cm]{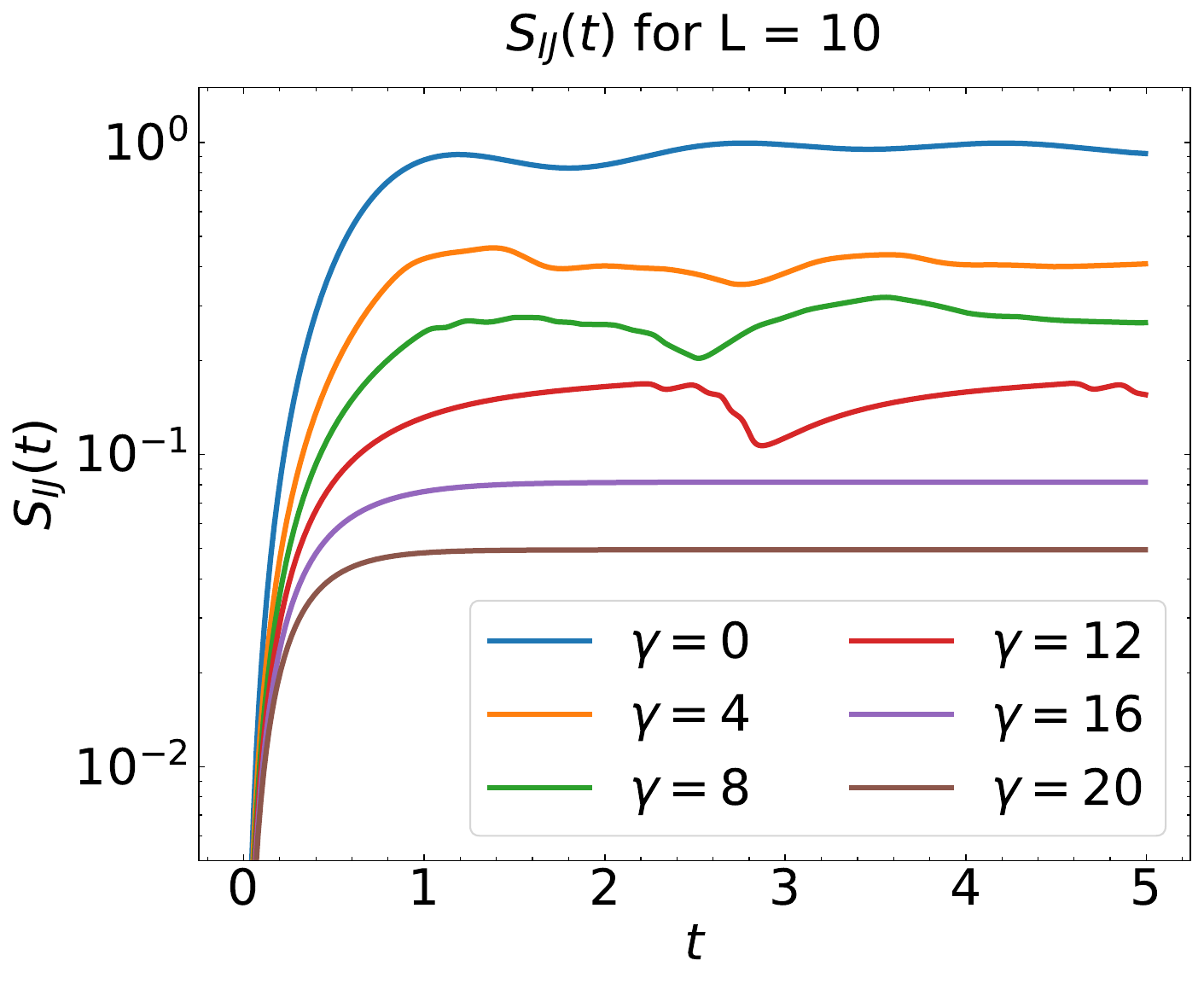}}
    \subfloat[]{\includegraphics[width=7cm]{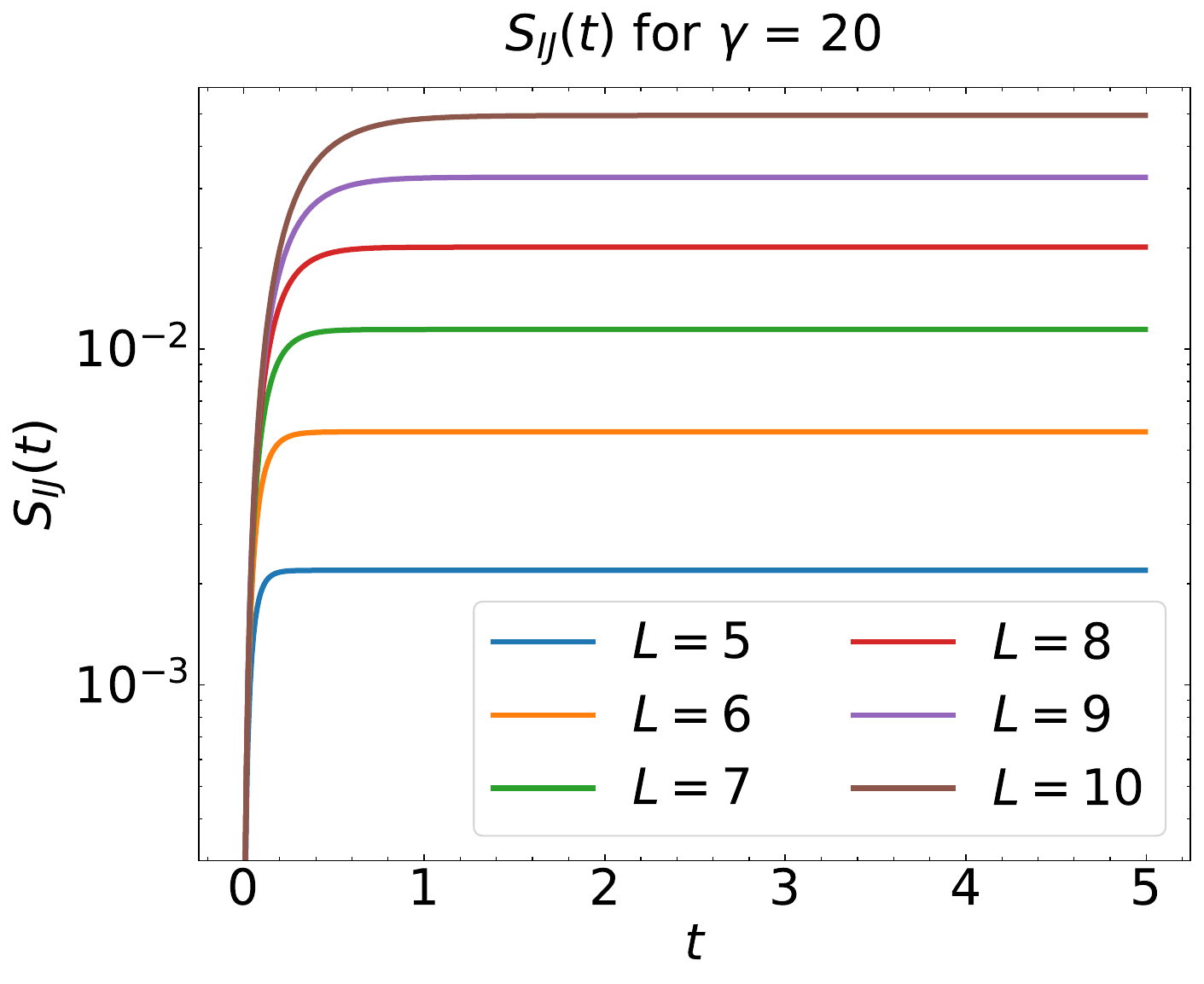}}
    \caption{
    Same as \figref{fig:asymptote}, but with a longitudinal field $h_\text{z}=1$.
    }
    \label{fig:asymptote_hz1}
\end{figure}

\begin{figure}
    \centering
    \subfloat[]{\includegraphics[width=7cm]{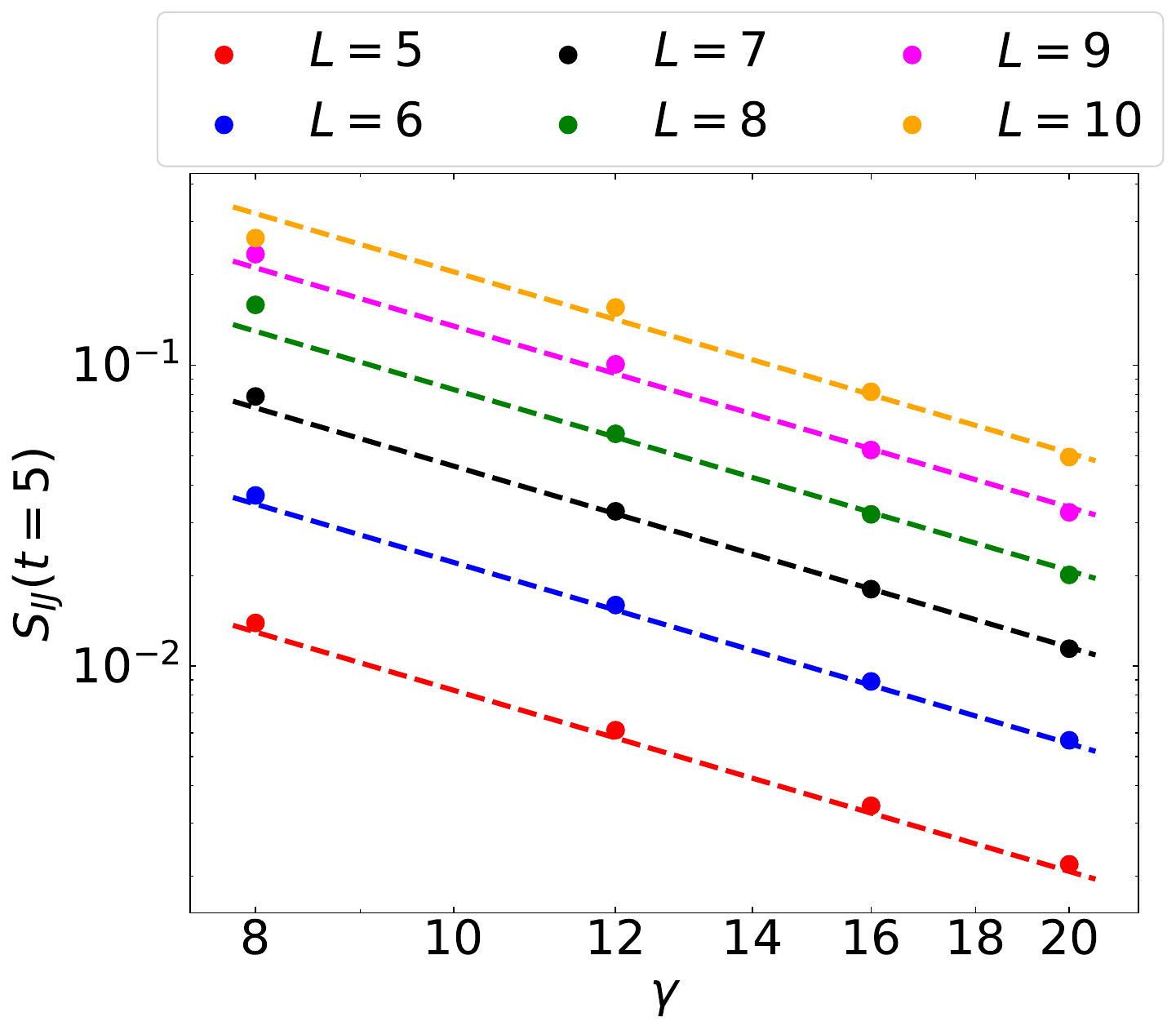}}
    \subfloat[]{\includegraphics[width=7cm]{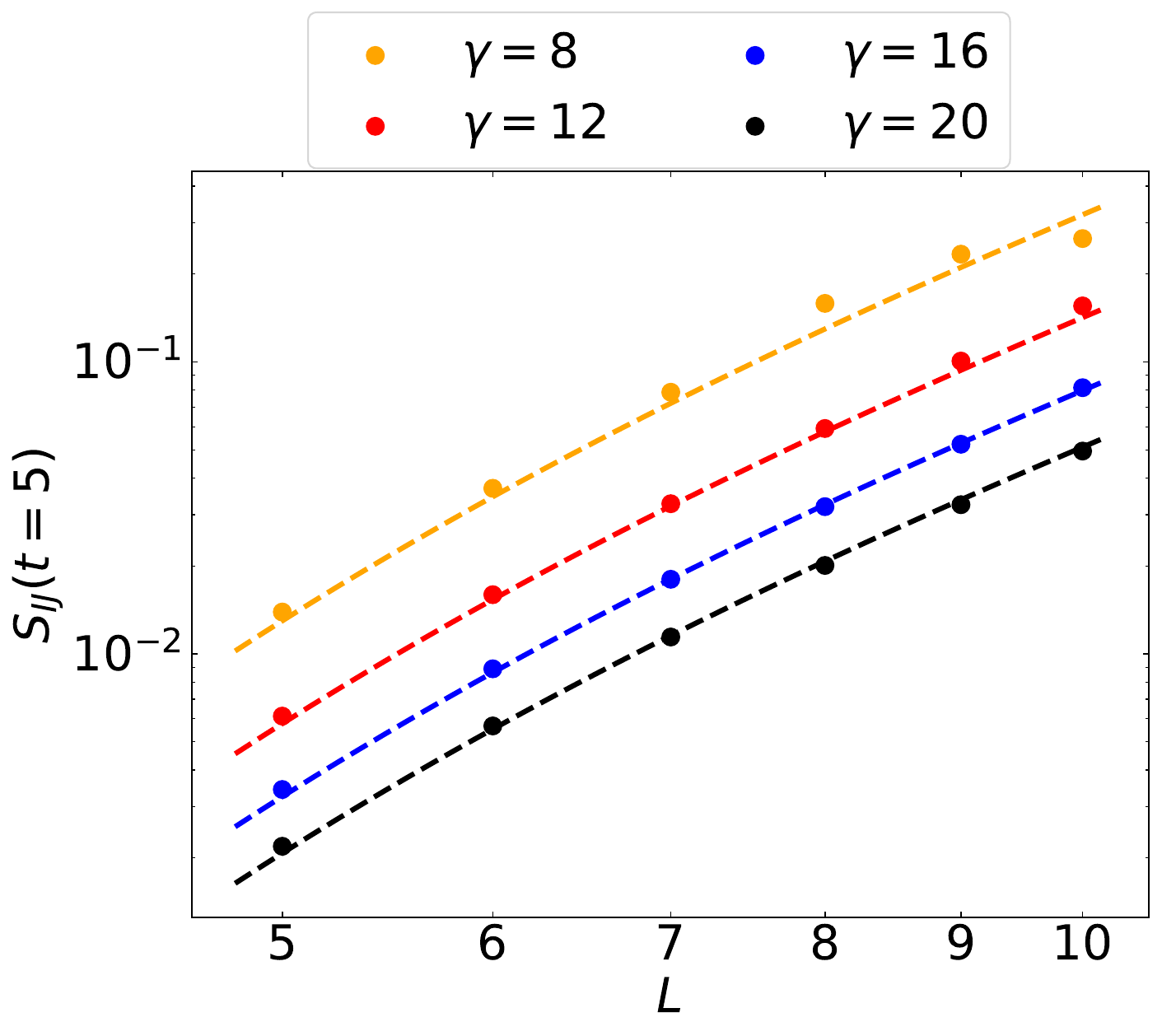}}
    \caption{Same as \figref{fig:scaling}, but with a longitudinal field $h_\text{z}=1$.}
    \label{fig:scaling_hz1}
\end{figure}

To demonstrate that the above qualitative results are generic,
  we also consider a non-integrable transverse field Ising model \cite{PhysRevB.92.104306} with an applied longitudinal field where $J=h_\text{x}=h_\text{z}=1$ in \eqnref{eq:hamil}.
\figsref{fig:asymptote_hz1} and \ref{fig:scaling_hz1}
  are analogous to \figsref{fig:asymptote} and \ref{fig:scaling}
    from the previous subsection.
We again find that the asymptote scales in accordance with \eqnref{eq:fit_asymp}.
We extract the fit parameters
\begin{align}
  a=4.21 \hspace{5mm} b=0.13 \hspace{5mm}c=-25.35
  \label{eq:param2}
\end{align}
  by fitting to \eqnref{eq:fit_asymp} using the 11 points given by $\gamma = 12,16,20$ and $L=7,8,9,10$ but with the $(\gamma=12, L=10)$ point excluded (which does not asymptote cleanly, as seen in \figref{fig:asymptote_hz1}).

\section{\texorpdfstring{$S_{IJ}(t)$ Squiggles Only for Small $\gamma$}{S(t) Squiggles Only for Small gamma}}
\label{sec:app_a}

\begin{figure}
    \centering
    \subfloat[\label{fig:sq_analysis1}]{\includegraphics[width=7cm]{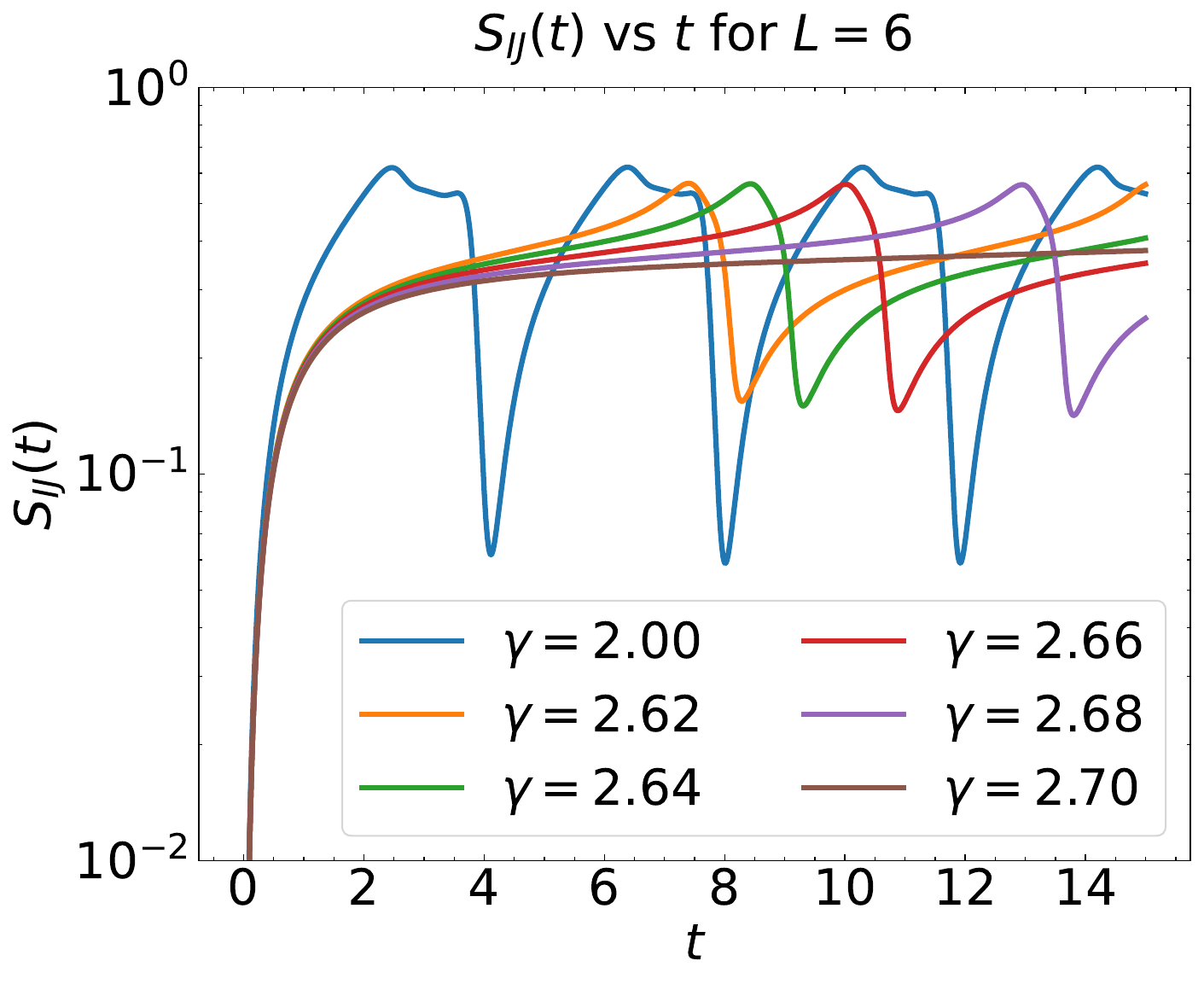}}
    \subfloat[\label{fig:sq_analysis2}]{\includegraphics[width=7.4cm]{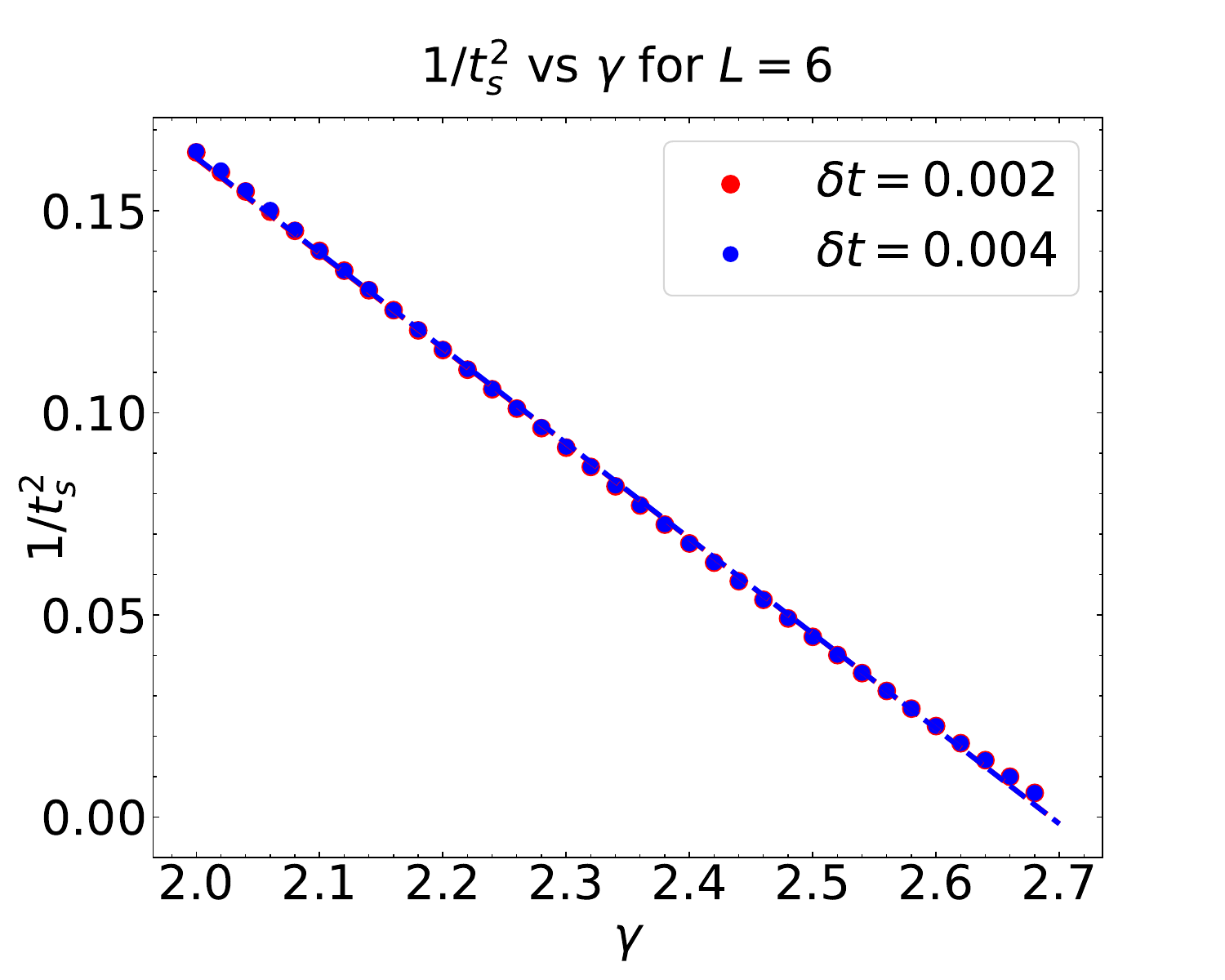}}
    \caption{
    {\bf (a)} Plot of $S_{IJ}(t)$ for $L=6$.
    As $\gamma$ approaches around 2.7,
      the temporal onset $t_s$ of the squiggles increases rapidly.
    {\bf (b)} 
    $1/t_s^2$ vs $\gamma$,
      where $t_s$ is the time at which the squiggles begin,
      which we take to be the time at which $S_{IJ}(t)$ first begins to decrease.
    The close fit to a straight line
      demonstrates the validity of \eqnref{eq:fit_sq} for large $t_s$ and shows that $t_s$ diverges at a finite $\gamma \sim 2.7$.}
    \label{fig:sq_analysis}
\end{figure}

In \figsref{fig:asymptoteL10} and \ref{fig:asymptoteL10_hz1},
  we observed that for large $\gamma$,
  $S_{IJ}(t)$ displayed a remarkably clean asymptote to a constant.
However for smaller $\gamma$,
  $S_{IJ}(t)$ instead wiggled around an approximate constant.
But the time at which the wiggles began appeared to increase with $\gamma$.
In this appendix,
  we show that the time at which the wiggles begin actually becomes infinite at a finite $\gamma$,
  which is strong evidence that there are no wiggles for sufficiently large $\gamma$.

Similar to \figref{fig:asymptoteL10}, in \figref{fig:sq_analysis1}
  we plot $S_{IJ}(t)$ for the critical transverse field Ising model ($h_{\text{z}}=0$), but for longer times and a smaller range of $\gamma$.
For computational efficiency, we first study a smaller system size of $L = 6$
  (with a Runge Kutta time step of $\delta t = 0.004$).
We clearly see that the onset of the squiggles increases rapidly as $\gamma$ approaches around 2.7.

Let $t_s$ be the time at which the squiggles begin,
  which we define as the time at which $S_{IJ}(t)$ first begins to decrease.
We find that $t_s$ diverges as
\begin{equation}
    t_s^2 = \frac{t_0^2}{\gamma - \gamma_0}
    \label{eq:fit_sq}
\end{equation}
  for large $t_s$
  where $\gamma_0$ and $t_0$ are constants.
To show this divergence, we rewrite the above equation as a linear equation in $t_s^{-2}$ vs $\gamma$:
\begin{equation}
     t_0^{2} \, t_s^{-2} = \gamma - \gamma_0
\end{equation}
In \figref{fig:sq_analysis2},
  we plot a series of points $(\gamma, t_s^{-2})$.
The points agree very precisely with a linear fit to the above equation,
  which is strong evidence that the temporal squiggle onset $t_s$
  diverges at $\gamma_0 \approx 2.7$.
We also show data points for numerical integration time step
  $\delta t=0.002$ (in addition to $\delta t=0.004$)
  to show that decreasing the time step by a factor of two has no noticeable affects on the data,
  which is good evidence that numerical integration errors are negligible for the data shown.
See \appref{sec:app_b} for additional details on numerical integration errors.

\begin{figure}
    \centering
    \subfloat[\label{fig:sq_analysisL8_1}]{\includegraphics[width=7cm]{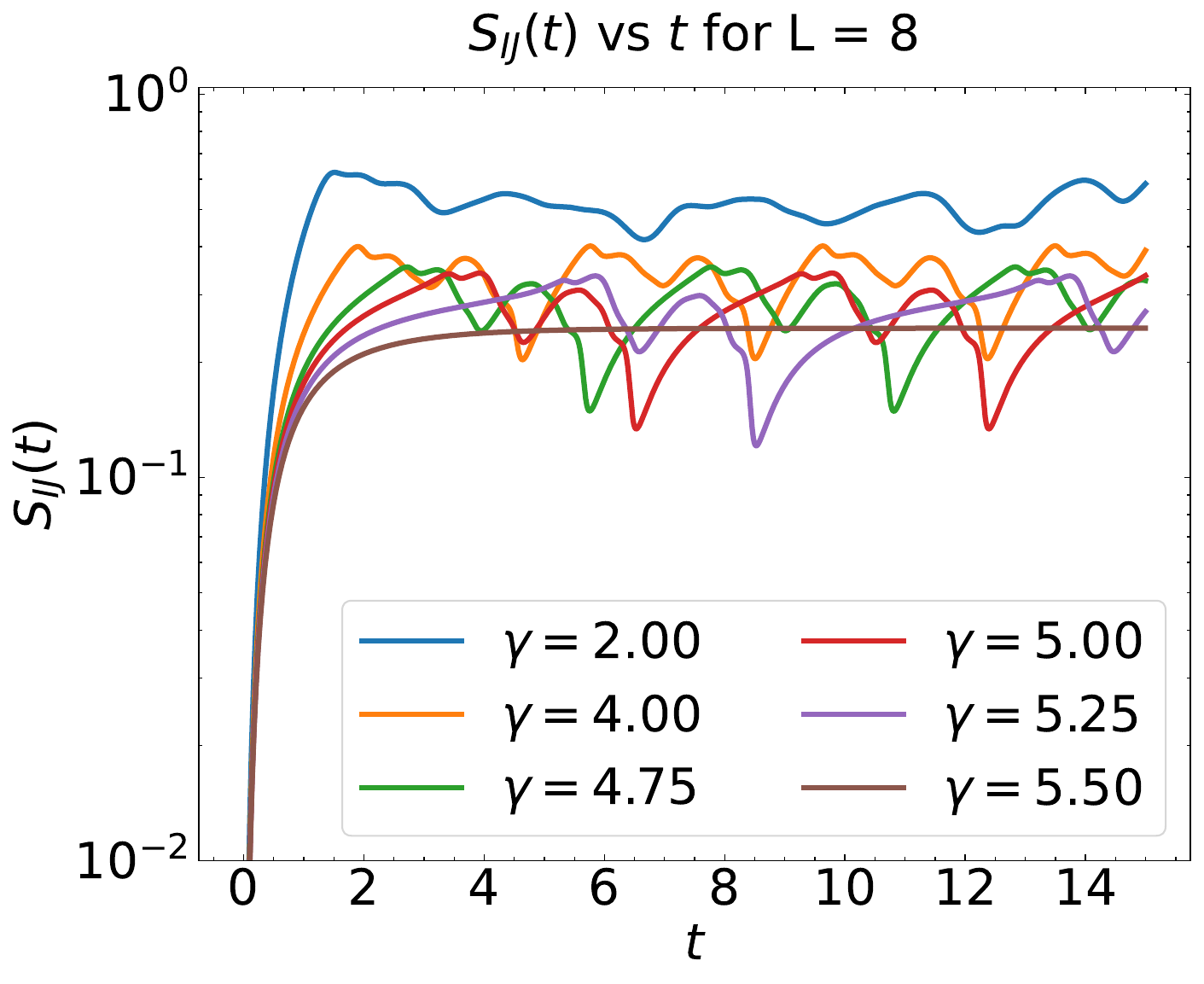}}
    \subfloat[\label{fig:sq_analysisL8_2}]{\includegraphics[width=7.4cm]{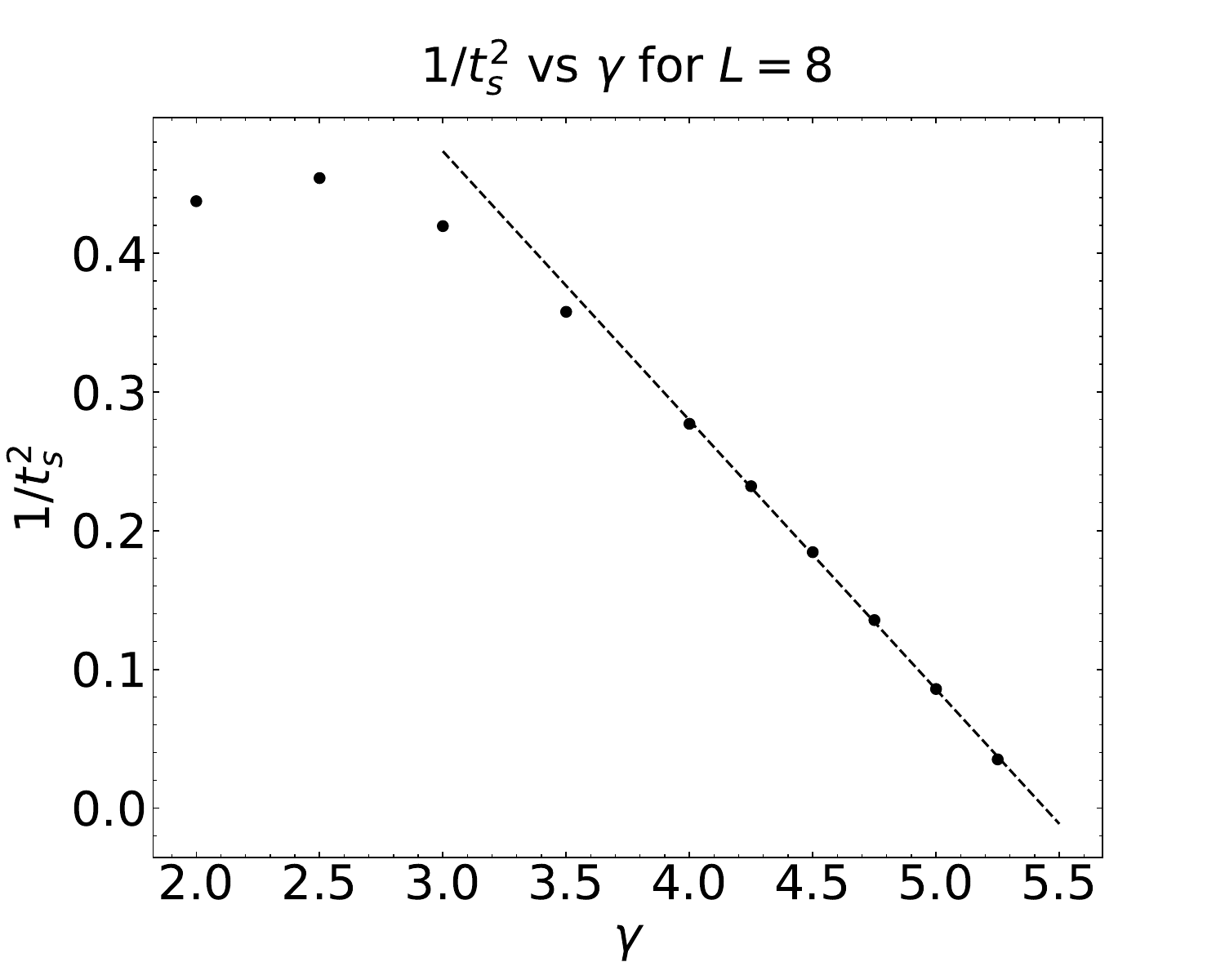}}
    \caption{
    Same as \figref{fig:sq_analysis} but for $L=8$ (and only $\delta t=0.004$).}
    \label{fig:sq_analysisL8}
\end{figure}

\begin{figure}
    \centering
    \subfloat[\label{fig:sq_analysis_ex1}]{\includegraphics[width=7cm]{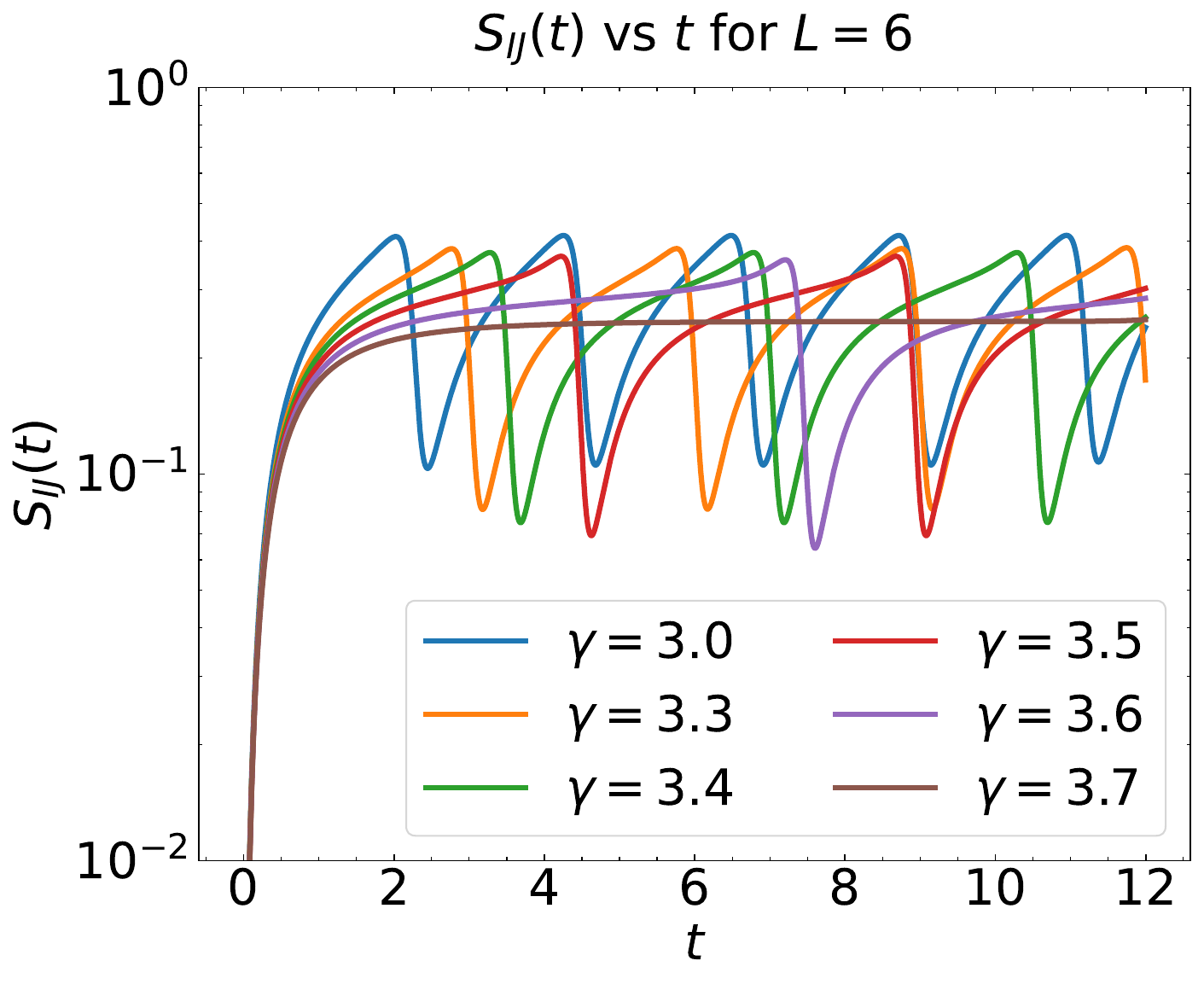}}
    \subfloat[\label{fig:sq_analysis_ex2}]{\includegraphics[width=7.4cm]{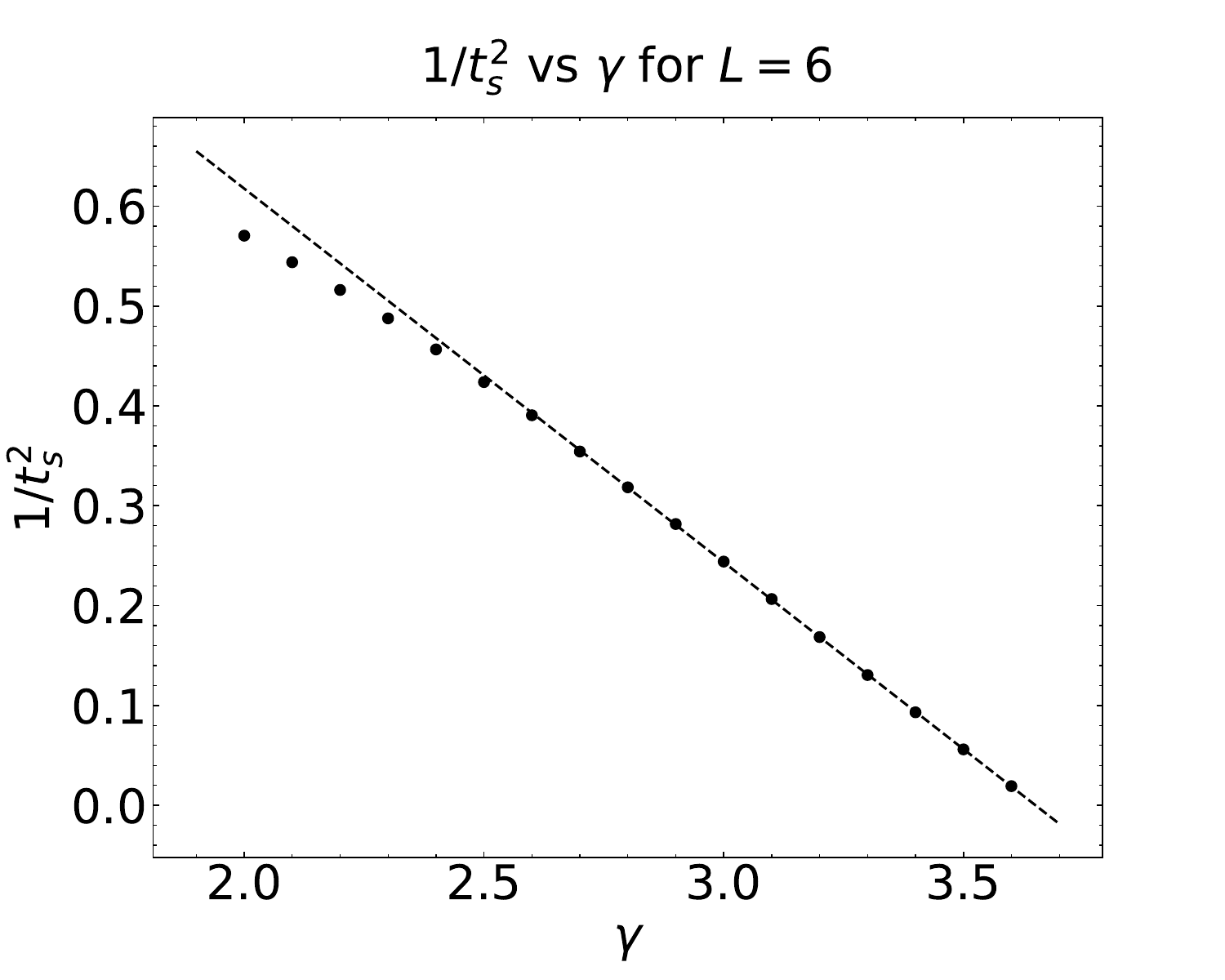}}
    \caption{Same as \figref{fig:sq_analysis} except but with an additional $h_{\text{z}} = 1$ longitudinal field (and only $\delta t = 0.004$).}
    \label{fig:sq_analysis_ex}
\end{figure}

To demonstrate the generality of this result,
  we also show data and analogous asymptotic fits for a larger system size $L=8$ in \figref{fig:sq_analysisL8},
  and with an applied $h_{\text{z}} = 1$ longitudinal field for $L=6$ in \figref{fig:sq_analysis_ex}. 
From \figref{fig:sq_analysisL8}, we see that the squiggles seem to persist up to larger $\gamma$ when the system size is larger.

\section{Conclusion}
\label{sec:conclusion}

Our work extends recent work on the gauge picture of quantum dynamics \cite{slagle2022gauge} to show that it is possible to modify the gauge picture such that the local wavefunctions in the gauge picture are approximately equal to the Schr\"odinger picture wavefunction
  while still enforcing equations of motion that are explicitly local
  (while Schr\"odinger's equation is not explicitly local).
This approximate equivalence of wavefunctions occurs when the connections in the gauge picture are close to the identity,
  which we obtain by adding an additional $\gamma$ term [see \eqsref{eq:G_int}, \eqref{eq:tildeX}, and \eqref{eq:X_I}] to the gauge picture equations of motion \eqref{eq:gaugeEoMG}.
We quantified how close the connections are to the identity via $S_{IJ}(t)$ in \eqnref{eq:Sch-ness}.
$S_{IJ}(t)$ therefore measures how ``close'' this modified gauge picture is to the Schr\"odinger picture,
  and $\gamma$ (in some sense) interpolates between the Schr\"odginer picture and (unmodified) gauge picture.
We showed that $S_{IJ}(t) \sim \gamma^{-2} e^{a L + b + \cdots}$ [\eqnref{sec:scaling}] 
for large $t$, large $\gamma$, and large system size $L$ using 1D spin chain numerics
  for the transverse-field Ising models that we studied.
(We expect that these numerical results apply to generic spin chain models.)
Thus, $\gamma$ must be exponentially large in the system size in order for the local wavefunctions in the gauge picture to be approximately equal to the Schr\"odinger picture wavefunction. 
We analytically argued that $S_{IJ}(t) \sim \gamma^{-2}$ in \secref{sec:scaling}.
We leave to future study: (1) a more rigorous analytical derivation, and (2)   an explanation for the system size dependence.

Interestingly, $S_{IJ}(t)$ exhibits an extremely clean asymptote as $t \to \infty$ for sufficiently large $\gamma$,
  but squiggles erratically for smaller $\gamma$.
We were surprised to find in \secref{sec:app_a} that there appears to be a sharp transition between these two regimes.
That is, for small $\gamma$ we find that $S_{IJ}(t)$ starts to show a clean asymptote at small $t$, but then begins to squiggle erratically at later times $t \gtrsim t_s$.
But we find that the onset time $t_s$ of the squiggles diverges at a finite $\gamma$.
It would be interesting to gain a better understanding of the physics underlying this effect.

We thoroughly checked that our data is free from numerical integration errors.
During that process, we found that the numerical integration errors are also interesting because they exhibit exponential sensitivity to initial conditions---the butterfly effect---even though quantum dynamics are linear.
However, the gauge picture equations of motion are nonlinear,
  and we expect that numerical errors turn on the nonlinearity,
  which we find leads to numerical errors that increase exponentially
  (which does not occur when integrating Schr\"odinger's equation).
See \appref{sec:app_b} for details.
Understanding this new kind of gauge picture chaotic dynamics would be another interesting direction for future research.

\section*{Acknowledgements}
We would like to thank Toni Panzera and Pranav Satheesh for useful conversations.

\paragraph{Funding information}
This research was supported in part by the Welch Foundation through Grant No. C-2166-20230405,
  and by the National Science Foundation Grant No. NSF PHY-1748958 and the Gordon and Betty Moore Foundation Grant No. 2919.02. 

\appendix

\section{Numerical Instability in the Gauge Picture}
\label{sec:app_b}

In this appendix, we study the numerical integration errors in some detail.
Nonlinear differential equations exhibit chaotic dynamics \cite{Li1975,lorenz1963deterministic,may1976simple,feigenbaum1978quantitative}.
That is, small perturbations to the initial conditions quickly evolve into large effects---the so-called butterfly effect.
Numerically integrating nonlinear differential equations therefore presents a challenge because small numerical errors are likely to similarly lead to large changes due to the same butterfly effect;
  but in this context these large changes are viewed as large integration errors.
Schr\"odinger's equation is a linear differential equation
  and therefore does not exhibit this kind of chaotic dynamics;
  small changes in $\ket{\psi(0)}$ does not lead to large changes in $\ket{\psi(t)}$ due to unitarity.
However, although the gauge picture equations of motion can reproduce the same expectation values as Schr\"odinger's picture when integrated exactly,
  in this appendix we find that small integration errors lead to exponentially increasing errors.
We expect this occurs because the small integration errors
  insert nonlinearity into the dynamics,
  which then exhibit the butterfly effect.

\begin{figure}
    \centering
    \includegraphics[width=15cm]{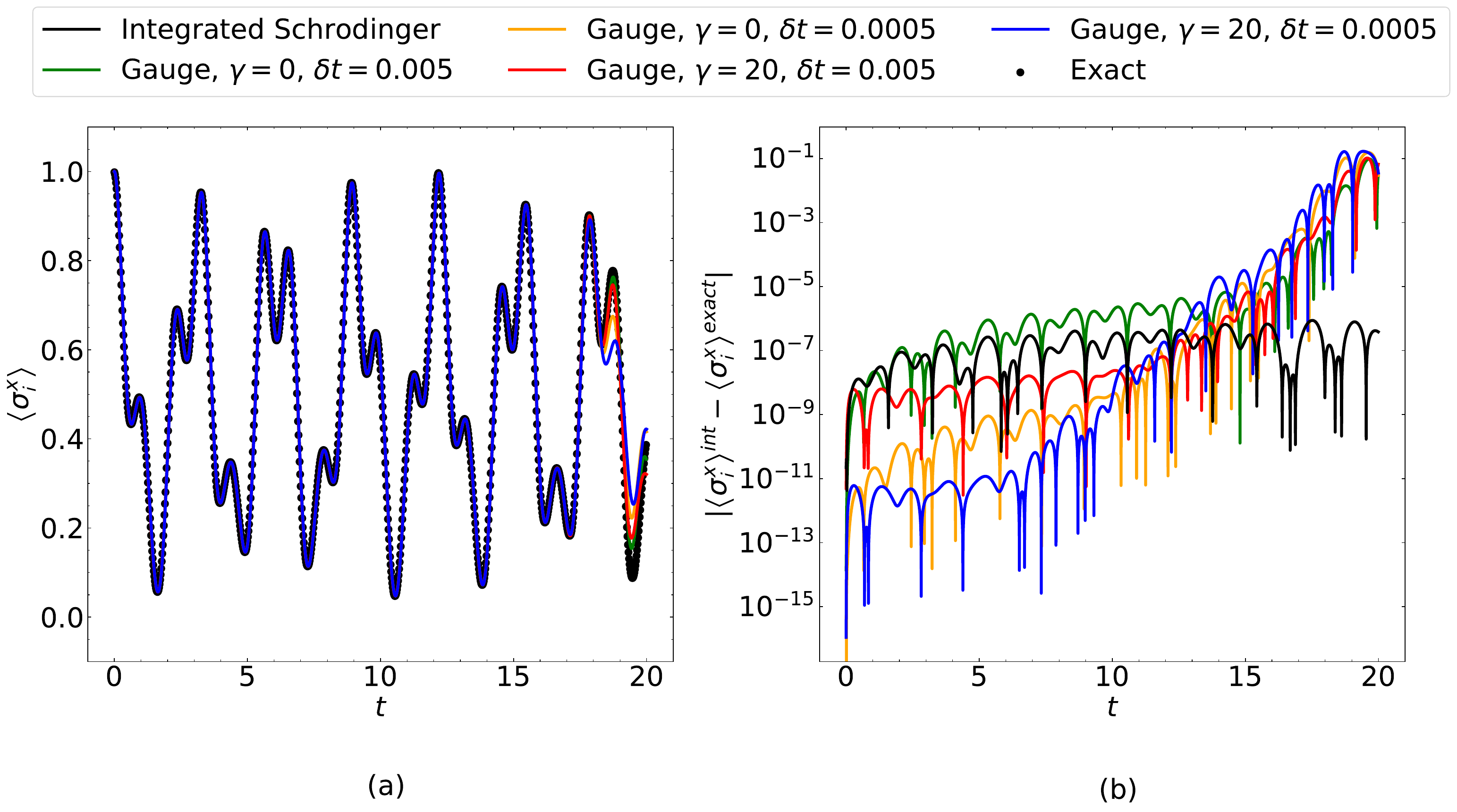}
    \caption{
    {\bf (a)} $\left\langle \sigma^{\text{x}}\right\rangle$ vs $t$ for $L=6$ comparing our gauge picture numerics (colored lines) to the numerically exact value (black).
    Gauge picture data is shown for different integration time steps $\delta t = 0.005$ or $0.0005$ and $\gamma = 0$ or $20$.
    We see good agreement until around $t \approx 18$.
    {\bf (b)} The error of gauge picture numerics (colored lines) on a log scale.
    We see that the error increases exponentially quickly,
      which is common for chaotic non-linear differential equations.
    For comparison, we also plot the error from integrating the Schr\"odinger picture equation of motion (black line);
      this error does not increase exponentially because Schr\"odinger's equation is linear.}
    \label{fig:chaos}
\end{figure}

In \sfigref{fig:chaos}{a}, we plot the $\braket{\sigma^x_i}$ expectation value of the critical $L=6$ transverse field Ising model (with $h_z=0$) vs time.
We compare our gauge picture numerics
  (which use the modified RK4 Runge Kutta integration method described in Appendix F of \refcite{slagle2022quantum})
  to our numerically exact Schr\"odinger picture numerics
  (for which we can simply exponentiate the Hamiltonian).
The two methods agree up until around time $t \approx 18$,
  at which point our gauge picture numerics become rather inaccurate.
We show gauge picture data for numerical integration times steps
  $\delta t = 0.005$ and $0.0005$.
In \sfigref{fig:chaos}{b}, we plot the error of the same expectation value on a log scale to show that the integration error increases exponentially with time.

\bibliography{interpolatingGaugePicture}

\nolinenumbers

\end{document}